\documentstyle[aps]{revtex}

\begin{document}
\twocolumn[\hsize\textwidth\columnwidth\hsize\csname
@twocolumnfalse\endcsname
\title{\bf Spectral noncommutative geometry 
and quantization: a simple example} 
\author{{\bf Carlo Rovelli}\\[.2cm]
{\em Centre de Physique Theorique, Luminy, F13288 Marseille, 
France}\\
{\em Physics Department, University of Pittsburgh, Pittsburgh, 
Pa 15260, USA}\\ 
rovelli@cpt.univ-mrs.fr} 
\date{\today}
\maketitle 
\begin{abstract}
We explore the relation between noncommutative geometry, in the 
spectral triple formulation, and quantum mechanics.  To this aim, 
we consider a dynamical theory of a noncommutative geometry 
defined by a spectral triple, and study its quantization.  In 
particular, we consider a simple model based on a finite 
dimensional spectral triple $(A, H, D)$, which mimics certain 
aspects of the spectral formulation of general relativity.  We 
find the physical phase space, $\Gamma$, which is the space of 
the onshell Dirac operators compatible with $A$ and $H$.  We 
define a natural symplectic structure over $\Gamma$ and construct 
the corresponding quantum theory using a covariant canonical 
quantization approach.  We show that the Connes distance between 
certain two states over the algebra $A$ (two ``spacetime 
points''), which is an arbitrary positive number in the classical 
noncommutative geometry, turns out to be discrete in the quantum 
theory, and we compute its spectrum.  The quantum states of the 
noncommutative geometry form a Hilbert space $K$.  $D$ is 
promoted to an operator $\hat D$ on ${\cal H}=H\otimes K$.  The 
triple $(A, {\cal H},{\hat D})$ can be viewed as the quantization 
of the family of the triples $(A, H, D)$.
\end{abstract}
\vskip.7cm
] 

\section{Why quantizing a noncommutative geometry} 

The idea that the geometric structure of physical spacetime could 
be noncommutative exists in different versions.  In some of 
versions, the noncommutativity of geometry is viewed as a direct 
effect of quantum mechanics, which disappears in the limit in 
which we consider processes involving actions much larger than 
the Planck constant \cite{noncaltri}.  In the noncommutative 
geometry approach of Connes et.\,al.\,(NCG) 
\cite{dirac,stmodel,sptr,thomas}, on the other hand, 
noncommutativity is introduced as a feature of spacetime which 
exists independently from quantum mechanics.  For instance, in 
the noncommutative version of the standard model \cite{stmodel}, 
the theory is defined over a noncommutative spacetime, 
and is {\it then\/} quantized along conventional perturbative 
lines.  More ambitiously, the spectral triple formulation 
\cite{sptr} includes the gravitational field as well.  For the 
gravitational field, however, conventional perturbative 
quantization methods fail \cite{qg}.  The problem of going from 
the noncommutative, but non-quantum-mechanical, spectral triple 
dynamics to the full quantum dynamics is thus open.

In a quantum theory that includes gravity, the geometric 
structure of spacetime is to be treated quantum mechanically.  
Therefore the noncommutative geometry of spacetime must be 
reinterpreted in quantum terms.  Thus, in a theory of the 
physical world based on NCG and including quantum mechanics, the 
geometry of spacetime should be represented by a {\it 
quantization\/} of a noncommutative geometry.

There should therefore be two distinct sources of 
noncommutativity in the theory: the noncommutativity of the 
elements of the algebra describing spacetime {\it and}, 
separately, the noncommutativity of the quantum mechanical 
variables.  In this note, we address the problem of understanding 
what a quantization of a noncommutative geometry might be, and 
what is the relation between geometric noncommutativity and 
quantum noncommutativity.  We study this problem using a simple 
model derived from \cite{thomas} with the aim of developing 
structures and notions which, hopefully, could guide us in 
addressing the same issue in a full model including general 
relativity.

In particular, we consider the spectral triple approach given in 
\cite{sptr}.  Within this approach, a dynamical model is given by 
the spectral triple $(A,H,D)$, where $H$ is a Hilbert space and 
$A$ is a $C^{*}$ algebra represented on $H$, which are fixed 
once and for all; while $D$ is a Dirac operator (in the sense of 
\cite{dirac}) in $H$, which codes the value of the dynamical 
fields, and in particular of the spacetime metric, that is, the 
gravitational field.  Thus $D$ is the dynamical variable of the 
model and represents a classical configuration of the theory.  
The dynamics is then given by an action $S[D]$.  To quantize the 
theory, we must find the Hilbert space $K$ of its quantum states.  
A state in $K$ will represent a quantum state of the 
noncommutative geometry: roughly, a probabilistic quantum 
superpositions of (noncommutative) geometries.  Such a state will 
assign not a number, but rather a probability distribution, to 
the observable distance $d(p,p')$ between any two points.  
Observable quantities will be represented by operators on $K$.

We construct the Hilbert space $K$ and the dynamical operators 
for our simple model.  From the quantum theory we obtain a 
concrete result: the physical distance between (certain) two 
points of the model, which in the classical theory can be an 
arbitrary nonnegative number $d$, turns out to be quantized as
\begin{equation}
	d = {\frac{ L_{P}}{\sqrt{2n+1}}}, \ \ \ \ n=0,1,2,3\ldots. 
	\label{qd}
\end{equation} 
where $L_{P}$ (the ``Planck length'') is the length determined by 
$\hbar$ and the coupling constant of the theory.  Furthermore, we 
show that the quantum theory can be compactly represented in 
terms of a novel triple $(A,{\cal H},\hat D)$, where $\cal H$ is 
the tensor product of $H$ with the space of the quantum 
geometries $K$.

There exist several attempts to explore the relation between 
noncommutative geometry and quantum theory by studying quantum 
fields defined over a noncommutative geometry 
\cite{noncaltri,ncgqm}.  These attempts should not be confused 
with the present work.  Here we are not concerned with the effect 
of a noncommutative structure of spacetime over quantum fields: 
we are concerned with the quantum mechanical properties of the 
noncommutative geometry itself.

\section{Preliminaries} 

\subsection{Mechanics without preferred time}

We begin by recalling a few simple points about classical and 
quantum mechanics.  This will fix notation and provide an 
appropriate conceptual framework.  We need to choose a language 
sufficiently general to deal with theories, such as general 
relativity (GR), in which time evolution enters in a non-simple 
manner.

A dynamical theory is defined by a set of equations, the 
equations of motion, for a set of variables, the dynamical 
variables.  By dynamical variable we mean here the full 
``history'', or ``motion'', of the physical system.  We denote 
the space of these variables as $\cal C$, or extended 
configuration space.  For instance, the dynamics of an oscillator 
is given by the equation of motion $d^{2}x(t)/dt^{2}=-\omega 
x(t)$, where the variable is the {\em function\/} $x: t \mapsto 
x(t)$.  Thus, in this example the extended configuration space 
$\cal C$ is $C_{\infty}(R)$.  When time evolution is standard, as 
in this example, $\cal C$ is the space of the maps from $R$ (the 
time) into a configuration space.  However, we are interested 
also in systems in which $\cal C$ does not have this form.  We 
consider Lagrangian systems, in which the equations of motion can 
be expressed as the vanishing of the first variation of a 
function $S[x]$ on $\cal C$.  $S[x]$ is the action functional.

The solutions of the equations of motion form a subspace of $\cal 
C$.  This subspace is the phase space of the system, and we 
denote it $\Gamma$.  In the oscillator, the solutions of the 
equations of motion are $x(t)=A \sin(\omega t+\phi)$.  Thus, 
$\Gamma$ is a two-dimensional subspace of $\cal C$ coordinatized 
by $A$ and $\phi$.\footnote{
In the example of the oscillator, and anytime time evolution is 
standard, we can represent $\Gamma$ as a space of initial 
data.  In fact, each set of initial data determines a solution 
and viceversa.  To do this, 
we have to fix a time, say $t=t_{0}$.  Define then $q=x(t_{0})$ 
and $p=dx(t)/dt|_{t=t_{0}}$.  The points in $\Gamma$ can then be 
coordinatized by $(p,q)$ instead of $(A,\phi)$.  The relation 
between the two set of coordinates on $\Gamma$ is immediately 
obtained from the definitions as $q=A \sin(\omega t_{0}+\phi)$ 
and $p=\omega A \cos(\omega t_{0}+\phi)$.  The $(p,q)$ 
coordinatization of $\Gamma$ is the one commonly introduced in 
textbooks.  The definition of the phase space as the space of the 
solutions of the equations of motion (known since 
Lagrange) has the advantages of being more covariant (this 
becomes clear in field theory), of not requiring the fixing of a 
preferred time $t_{0}$ and of being extendible to the dynamical 
systems without standard time evolution that concern us here.}

A point $s$ in $\Gamma$ represents a physically realizable (that 
is, compatible with the classical equations of motion), or 
``onshell'', history of the system.  This is a classical state of 
the system.  Here, ``state'' is used in the same (atemporal) 
sense as ``Heisenberg state'' in quantum theory.  An observable 
quantity $f$ corresponds to a function $f: s\in\Gamma \mapsto 
f(s)\in R$, where $f(s)$ is the predicted value of $f$ in the 
state $s$.\footnote{For instance, the position $x_{t}$ of the 
oscillator at time $t$ is the function $x_{t}(A,\phi)= A 
\sin(\omega t+\phi)$ on $\Gamma$.} The phase space carries a 
symplectic structure.  In the example, $\Omega = dp\wedge dq = 
\omega A\, dA\wedge d\phi$.  This defines Poisson brackets 
between observables.  If the system has a standard time 
structure, the symplectic structure can be determined by standard 
methods.  The definition of $\Omega$ in more general cases can be 
problematic.  The Poisson bracket structures of the observables 
is the starting point for their quantization.

\subsection{The spectral formulation of general relativity}

The model we introduce in the next section is meant to mimic some 
of the features of the spectral formulation of GR 
\cite{sptr,landi}.  We briefly recall this formulation.  Consider 
euclidean GR over a fixed compact\footnote{Euclidean GR over a 
compact manifold is a theory which is likely to admit only a 
finite dimensional space of solutions, and whose relation with 
physical GR is questionable.  However, the theory is, by itself, 
relevant as a model of diffeomorphism invariant field theory of 
the geometry.  More importantly, the experience with conventional 
euclidean quantum field theory suggests that the quantization of 
the euclidean theory with compact time might be of great 
relevance for understanding the true quantum theory even if the 
corresponding classical theories have extremely different 
properties.} four dimensional manifold $M$.  Let ${\cal G}_{M}$ 
be the space of the riemanian metrics $g:M\times M \to R^{+}$ 
over $M$.  In the standard formulation of the theory, ${\cal 
G}_{M}$ is taken as the extended configuration space
\begin{equation}
 	 {\cal C} \equiv {\cal G}_{M}. 
     \label{spacegr}
\end{equation}
This is the space of the (four-dimensional) gravitational fields.  
The action is then chosen to be the the well known 
Einstein-Hilbert action
\begin{equation}
	S[g]=\int d^{4}x\  \sqrt{g}R 
	\label{actiongr1}
\end{equation}
The phase space $\Gamma$ is the space of the Einstein spaces, 
namely the gravitational fields that solve the Einstein 
equations.

The theory can be formulated also as follows.  Fix the 4d compact 
(spin) manifold $M$.  Consider the spectral triples $(A,H,D)$.  
Here $A=C_{\infty}(M)$; the Hilbert space $H$ is the (Hilbert 
completion of the) space of the half-densitized Dirac spinor 
fields on $M$.  Notice that $H$ is defined also in the absence of 
a Riemanian structure\footnote{I thank Alain Connes for 
clarifying this point.}.  $H$ carries a representation $\pi$ 
of $A$, where $\pi(a)$ is the operator that multiplies the spinor 
by the function $a\in A$.  $A$ and $H$ are fixed structures.  We 
then consider the space ${\cal D}_{(A,H)}$ of all the Dirac 
operators $D$, in the sense of Connes \cite{dirac}, namely the 
set of the operators $D$ on $H$ such that $(A,H,D)$ is a spectral 
triple.  We take the space of the Dirac operators as the extended 
configuration space
\begin{equation}
 	 {\cal C} \equiv {\cal D}_{(A,H)}.
     \label{space}
\end{equation}
Thus, a Dirac operator represents here a history of the 
gravitational field. The action is chosen to be
\begin{equation}
	S[D] = Tr\ [f(D)]
	\label{actiongr}
\end{equation}
where $f(\cdot)$ is a suitable simple function given in 
\cite{sptr,landi}.  The dynamical system defined by (\ref{space}) 
and (\ref{actiongr}) is physically equivalent to the one defined 
by (\ref{spacegr}) and (\ref{actiongr1}) (see \cite{sptr,landi} 
for qualifications).  The reason is that given a metric 
$g\in{\cal G}_{M}$, there exists a corresponding Dirac operator 
$D(g)$ on $H$, defined with standard techniques.  Viceversa, 
given an operator $D$ in ${\cal D}_{(A,H)}$ there is a Riemanian 
metric $g$ on $M$ defined by $g=d$, where $d$ is given by the NCG 
distance formula
\begin{equation} 
d(p,p') = sup_{\{a\in A, |[D,\pi(a)]|
{\scriptscriptstyle \le} 1\}}\, |p(a)-p'(a)|, 
\ \  p,p'\in M,   
\label{distance}
\end{equation}
which is such that $D=D(g)$.  In (\ref{distance}), $p(a)=a(p)$ 
and $p'(a)=a(p')$ are the Gel'fand states (the points of the 
space $M$ correspond to states over the algebra $C_{\infty}(M)$).  
Thus, there is a one-to-one correspondence between the space 
${\cal D}_{(A,H)}$ of the Dirac operators and the space ${\cal 
G}_{M}$ of the Riemanian metrics, and we can identify the two 
spaces.  Finally, it is shown in \cite{dirac} that $S[D(g)]\sim S[g]$ 
(again, see \cite{sptr,landi} for qualifications).

The key lesson we learn from this is that the space of the Dirac 
operators ${\cal D}_{(A,H)}$ can be viewed as the space $\cal C$ 
of the extended configurations of the gravitational field.  
Accordingly, the phase space $\Gamma$ of the theory is the 
subspace of ${\cal D}_{(A,H)}$ where the first variation of the 
action functional $S[D]$ vanishes.  In a (nonperturbative) 
quantization of GR, one chooses a coordinatization $f,f', \ldots 
$ of the phase space $\Gamma$ of the theory and then searches for 
a representation of the Poisson algebra of the observables $f,f', 
\ldots $ in terms of self-adjoint operators on a Hilbert space 
$K$.  If we want to explore non-perturbatively the quantum 
mechanics of a NCG model, we have thus to quantize a phase space 
$\Gamma$ formed by the Dirac operator obeying appropriate 
equations of motion.  In other words, we have to coordinatize 
this space, and represent the coordinates as operators on the 
quantum state space.  Below, we complete this procedure for a 
simple, finite dimensional, model.

\section{Definition of the model} 

We choose a finite dimensional spectral triple $(A,H,D)$ (see 
\cite{thomas}).  Let $A$ be the algebra $M_{2}\oplus C$, where 
$M_{2}$ is the algebra of complex $2\times 2$ matrices and $C$ is 
the complex plane.  We write $a=({\mathrm{A}},\alpha)\in A = 
M_{2}\oplus C$.  Let $H$ be the Hilbert space $M_{3}$, the linear 
space of complex $3\times 3$ matrices $\Psi$ with the scalar 
product $(\Psi,\Phi)=Tr[\Psi^{\dagger}\Phi]$, and let $A$ act on 
$H$ in the representation $\pi$ given by
\begin{equation}
 	\pi(a) \Psi = \Psi_{a}\Psi
\end{equation} 
where 	
\begin{equation}
 	\Psi_{a} = 
 	\pmatrix{{\mathrm{A}} & 0 \cr 0 & \alpha}. 
	\label{f} 
\end{equation} 
A Dirac operator on $H$ \cite{dirac,thomas}, has the form (see 
\cite{thomas})
\begin{equation} 
	D\Psi= {{\mathrm{D}}}\Psi	+ \Psi {{\mathrm{D}}}^{\dagger}
\end{equation} 
where 
\begin{equation}
	{{\mathrm{D}}}=\pmatrix{0 & m \cr \overline m & 0} = 
	\pmatrix{0 & 0 & m_{1} \cr 0 & 0 & m_{2} \cr \overline m_{1} & 
	\overline m_{2} & 0 }.
	\label{dirac}
\end{equation}
$m_{1}$ and $m_{2}$ are complex numbers.  As mentioned 
above, we view $A$ and $H$ as fixed structures, while $D$ is the 
dynamical variable of the model.  Since the Dirac operator 
(\ref{dirac}) is determined by the two complex numbers $m_{i}, 
i=1,2$, the space ${\cal D}_{(M_{2}\oplus C,C^{3})}$ of the Dirac 
operators compatible with the given $A$ and $H$, namely the space 
of the dynamical variables, is isomorphic to $C^{2}$ and 
coordinatized by $m_{i}$.  Thus the extended configurations space 
is
  \begin{equation}
  	{\cal C} = {\cal D}_{(M_{2}\oplus C,\,M_{3})} \sim C^{2}. 
  \end{equation}  
This is the analog, in our model, of the space of the 
configurations of the gravitational field in GR

We complete the definition of the spectral triple (see 
\cite{dirac}) by defining 
\begin{eqnarray}
    J\Psi &\equiv& \Psi^{\dagger}; 	
\nonumber \\
    \Psi_{\gamma} &\equiv& \pmatrix{1&0\cr 0&-1}, 
   \nonumber \\
    \gamma \Psi &\equiv& \Psi_{\gamma}\Psi, 
\nonumber \\
    \chi\Psi &\equiv& \gamma J\gamma J \Psi = 
    \Psi_{\gamma}\Psi\Psi_{\gamma}.
\end{eqnarray}
The two involutions $J$ (``charge conjugation'') and $\chi$ 
(``parity'') satisfy the properties needed to complete the 
definition of the spectral triple.

The dynamics is determined by choosing an action $S[D]$, namely a 
function on the space of the dynamical variables $\cal C$.  Here 
we disregard the ``spectral principle'' \cite{sptr} which 
requires the action to depend on the spectrum of $D$ only: we 
leave the extension of the present ideas to genuinely spectral 
invariant actions to further developments.  The simplest 
possibility is to have a ``free'' theory with a quadratic action.  
Thus we search an action of the form
\begin{equation}
	S[D] = \ \frac{1}{2}\ Tr[\,{\mathrm{D}}\,\tilde M\,{\mathrm{D}}\,],
	\label{action}
\end{equation}
where $\tilde M$ is a $3\times 3$ matrix that determines the 
equations of motion.  It is easy to see that, because of the 
special form (\ref{dirac}) of ${\mathrm{D}}$, the action $S[D]$ in 
(\ref{action}) can be rewritten as
\begin{equation}
	S[D] = \ \overline m_{i} M_{ij} m_{j}, 
	\label{action2}
\end{equation}
where $M$ is a $2\times 2$ matrix.  We want the action to be 
real, and thus $M$ must be hermitian.  We want the theory to have 
a nontrivial space of solutions, and thus $M$ must have vanishing 
determinant.  These requirements fix $M$ up to a single complex 
parameter $\alpha$ and an overall normalization $G$ that does not 
affect the field equations but is needed to set the right 
physical dimensions
\begin{equation}
	M = \frac{1}{G} \pmatrix{|\alpha|^{2} & \overline\alpha \cr 
	\alpha & 1}.
	\label{M1}
\end{equation}
(We write $G$ in the denominator, following the GR use.)
For simplicity, we further choose $\alpha$ to be a pure phase 
$\alpha=e^{i\phi}$, although this is not really needed for what 
follows.   Thus
\begin{equation}
	M = \frac{1}{G} \pmatrix{ 1 & e^{-i\phi} \cr e^{i\phi} & 1}.
	\label{M}
\end{equation} 
This completes the definition of the model. Let us now 
analyze its content. 

By extremizing the action with respect to $m_{i}$ and $\bar 
m_{i}$ we obtain the equations of motion  
\begin{equation}
	m_{2}=  e^{i\phi} \ m_{1},  
	\label{eqofm}
\end{equation}
the ``Einstein equations'' of the model.  These select, out of 
all the Dirac operators, a physical phase space $\Gamma$ of 
physical ones (in GR: the Dirac operators corresponding to 
Einstein metrics). We say that $D$ is ``on shell'' if it 
satisfies the equations of motion. 

The phase space $\Gamma$ is coordinatized just by $m_{1}$.  
Therefore $\Gamma$ is isomorphic to the complex plane $C$.  From 
now on we will write $ m \equiv m_{1}$ for simplicity.  The 
complex plane has a natural symplectic structure. 
\begin{equation}
	\Omega = \frac{i}{G} \ dm \wedge d\overline m = 
	\frac{2}{G}\ d(\Re(m))\wedge d(\Im(m)).
	\label{omega}
\end{equation}
($G$ adjusts dimensions.)  We take this as the physical 
symplectic structure of the system\footnote{I thank Abhay 
Ashtekar for this suggestion}. Thus the basic Poisson brackets 
are 
\begin{equation}
	\{ m , \overline m \} = iG 
	\label{pb}
\end{equation} 

A straightforward computation shows that the symplectic form 
$\Omega$ can be written directly in terms of the Dirac operator. 
We define by $\Omega_{ex} \equiv d\theta$, where 
\begin{equation}
 \theta \equiv \frac{i}{12G}\ {\rm Tr}[\gamma D dD] = \frac{i}{4G} 
\  Tr[\Psi_\gamma\mathrm{D}d\mathrm{D}]. 
\end{equation} 
(The first trace is the trace of the operators on $H$, 
the second is the trace of the $3\times 3$ matrices.)
Then $\Omega$ is the restriction on shell of $\Omega_{ex}$. 

The physical interpretation of $D$ is that it determines the 
metric structure, and therefore the ``gravitational field'', over 
the space of the states over $A$. A state $p$ over $A$ is 
determined by a vector $\Psi_{p}$ in $H$
\begin{equation}
	p(a) = \langle \Psi_{p} |\pi(a)| \Psi_{p} \rangle. 
\end{equation}
In particular, consider the two vectors 
$\Psi_{p}=\psi_{p}\otimes\psi_{p}$ and 
$\Psi_{p'}=\psi_{p'}\otimes\psi_{p'}$, where $\psi_{p}=(0,0,1)$ 
and $\psi_{p'}=\frac{1}{\sqrt{2}}(e^{i\phi},1,0)$.  They 
define the states $p$ and $p'$ over $A$.  Their distance is given 
by (\ref{distance}) and can be explicitly computed using 
Eq.\,(2.175) of Sec\,2.4.1 in \cite{thomas}, obtaining
\begin{equation}
	d(p,p')=\frac{1}{\sqrt{|m_{1}|^{2}+|m_{2}|^{2}}}
	=\frac{1}{\sqrt{2}|m|}.  
	\label{d}
\end{equation} 

\section{Quantization} 

We want to promote $m$ and $\overline m$ to operators 
$\hat m$ and $\hat{\overline m}$ on a Hilbert space $K$ in such a 
way that their algebra represents ($i\hbar$ times) the Poisson 
algebra (\ref{pb}) and that the operators $\Re\hat m=1/2(\hat 
m+\hat{\overline m})$ and ${\Im}\hat m =i/2(\hat m-\hat{\overline 
m})$, which correspond to real quantities, be self-adjoint.  The 
solution is well known.  The complex structure of $\Gamma$ 
provides us with a preferred polarization: we choose a 
representation in which the wave functions depend on $m$ and are 
independent from $\overline m$.  Vectors in $K$ are thus analytic 
functions $\psi(m)$ and
\begin{eqnarray}
	\hat m\ \psi(m) &=& m\ \psi(m), \nonumber \\
	\hat{\overline m}\ \psi(m) &=& \hbar G\ \frac{d}{dm}\psi(m). 
	\label{mbarm}
\end{eqnarray} 
The scalar product that implements the desired reality conditions is 
\begin{equation}
(\psi,\psi') = \frac{1}{\pi\hbar G} \int_{C} 
dm \ e^{-\frac{|m|^{2}}{\hbar G}}\ 
\overline{\psi(m)} \, \psi'(m)
\label{sp}
\end{equation}
(The notation is $dm=d\Re(m) d\Im(m)$.)  A convenient orthonormal 
basis is given by the monomials
\begin{equation}  
\psi_{n}(m)=\frac{m^{n}}{\sqrt{n!}(\hbar G)^{n/2}}= \langle m | n \rangle,
\label{basis}
\end{equation}
where we have introduced the Dirac notation $| n \rangle$ for the 
basis elements.  In this basis $\hat m$ and $ \hat{\overline m}$, 
\begin{eqnarray}
	\hat m\ | n \rangle & = & \sqrt{\hbar G}\ \sqrt{n+1}\ | n+1 \rangle,
	\nonumber  \\ 
	\hat{\overline m}\ | n \rangle & = 
	& \sqrt{\hbar G}\ \sqrt{n}\ | n-1 \rangle, 
	\label{creann}
\end{eqnarray}
are immediately recognized as the creation and annihilation 
operators.  In fact, equations (\ref{mbarm},\ref{sp},\ref{basis}) 
form the standard definition of the harmonic oscillator quantum 
mechanics in the Bargmann representation.  We write the vectors 
in $K$ as
\begin{equation}
  	| \psi \rangle = \sum_{n} \ \psi_{n}\, |n\rangle,
\end{equation}
namely we represent $K$ as the $l_{2}$ space of sequences $\psi_{n}$. 

The operator corresponding to the $m_{2}$ variable is related to 
$\hat m_{1}=\hat m$ by the (Heisenberg) equations of motion
\begin{equation}
	\hat m_{2}= e^{i\phi}\ \hat m_{1}.
\label{hei}
\end{equation}

Consider now the distance  $d$ between the two points $p$ and $p'$.  
In the classical theory, this is given by (\ref{d}).  The 
corresponding operator $\hat d$ in the quantum theory is obtained 
by replacing the classical quantities $m_{i},{\overline m_{i}}$ 
with their quantum counterparts $\hat m_{i}, \hat{\overline 
m_{i}}$.  Using (\ref{hei}), we obtain
\begin{equation}
	\hat d = \frac{1}{\sqrt{2N}},
\end{equation}
where $N$ is the operator corresponding to the classical quantity 
$m\overline m$.  We encounter here an ordering ambiguity.  We compute  
easily from (\ref{creann})
\begin{equation} 
	N  | n \rangle = \hbar G \ (n+c)\  | n \rangle.
\end{equation}
where we have $c=0$ if we order $N$ as $N=\hat m\hat{\overline 
m}$; we have $c=1$ with the inverse ordering $N=\hat{\overline 
m}\hat m$; and we have the well known ``harmonic oscillator 
vacuum energy'' $c=1/2$ with the ``natural'' symmetric ordering, 
which we assume from now on. 

Thus we obtain the result that in the quantum theory the distance 
$d(p,p')$ is quantized, with discrete eigenvalues
\begin{equation}
d(p,p')  = \frac{1}{\sqrt{(2n+1)\hbar G}}
	=  \frac{L_{P}}{\sqrt{2n+1}}, 
\label{main}
\end{equation}
where 
\begin{equation}
       L_{P}=\frac{1}{\sqrt{\hbar G}}. 
       \label{LP}
\end{equation}
In the classical limit in which $\hbar$ is small and the quantum 
number $n$ is large,  the eigenvalues of the distance become more 
and more dense and approximate the classical continuum.

\section{Quantum noncommutative geometry} 

So far we have introduced two Hilbert spaces.  The first is the 
Hilbert space of the spectral triple we started from.  This is 
$H=M_{3}$, with vectors $\Psi$.  We write their components also 
as $\psi^{ab}, a,b=1,2,3$.  $H$ is the space of the ``fermions'' 
over the spectral triple.  The second is the Hilbert space of the 
quantum theory $K=l_{2}$, with vectors $\psi_{n}, 
n=0,1,2,\ldots$.  $K$ is the space of the quantum geometries.  
That is, a vector in $K$ can be seen as a quantum linear 
superposition of different classical geometries, or a quantum 
superposition of Dirac operators.  Each such states assigns not a 
fixed value, but rather a probability distribution over the 
possible values of the geometry defined by $D$.  Accordingly, in 
the quantum theory the Dirac operator (an operator on $H$) is 
promoted to an operator on $K$, or, more precisely to an object 
which is at the same time an operator on $H$ and on $K$:
\begin{equation}
	\hat D = \pmatrix{0 & 0 & \hat m_{1} \cr
	0 & 0 & \hat m_{2} \cr
	\hat{\overline m_{1}} & \hat{\overline  m_{2}}& 0} .
	\label{hD}
\end{equation}

It is thus natural to consider the tensor product of $H$ and $K$.  
We denote this tensor product as ${\cal H}=H\otimes K$.  The 
vectors in $\cal H$ can be written as $\psi_{n}^{ab}$.  In this 
basis, the matrix elements of the quantized Dirac operator $\hat 
D$ can be computed directly from (\ref{creann}) and (\ref{hD}). 
They are 
\begin{eqnarray}
	\hat D{}_n^{ab}{}^{m}_{cd} &=& \sqrt{\hbar G}\ \delta_{d}^{b} \ 
	\left[\sqrt{n}\ (\delta^{a}_{1}+\delta^{a}_{2}) 
	\ \delta_{c}^{3}\ \delta_{n}^{m-1}\right. \nonumber \\ &&
	\left. + \sqrt{n+1}\ \delta^{a}_{3}
	\ \delta_{n}^{m+1}\ (\delta_{c}^{1}+\delta_{c}^{2})\right] .
	\label{hatD}
\end{eqnarray}
Notice that this is a {\em single\/} operator, not a variable 
anymore (as the field operator in a quantum field theory is a 
single operator, not a variable as the classical field). 

The algebra $A$, as well as $J$ and $\chi$, are still represented 
in $\cal H$ (they act trivially on $K$).  Thus the quantum theory 
defines a novel triple $ (A, \ {\cal H},\ \hat D) $ with respect 
to which the set of the initial spectral triples $(A,H,D)$ (for 
all possible $D$'s), can be seen as a classical limit.  The 
representation of $(A, J, \chi)$ in $\cal H$ is highly reducible.  
Each state $\psi$ in $K$, namely each quantum geometry, defines a 
3d subspace of $\cal H$ carrying a irreducible faithful 
representation $\pi_{\psi}$.  We call $D_{\psi}$ the operator 
with matrix elements
\begin{equation}
	(D_{\psi})_{cd}^{ab} = \overline\psi^{n} \hat 
	D{}_n^{ab}{}^{m}_{cd}  
	\psi_{m}
\end{equation}
acting on this subspace.  Then the expectation value of the 
distance between any two states $p$ and $p'$ over $A$, in the quantum 
geometry determined by $\psi$ is 
\begin{equation}
	d_{\psi}(p,p') = 
	sup_{\{a\in A, |[D_{\psi},\pi_{\psi}(a)]|
	{\scriptscriptstyle \le} 1\}}\ |p(a)-p'(a)|, 
\end{equation}
where, notice, the Dirac operator and the representation of the 
algebra are restricted to the irreducible component.  Conversely, 
each irreducible representation of $(A, J, \chi)$ in $\cal 
H$ determines a state in $K$.  It is thus tempting to identify 
the quantum states of the geometry with these irreducible 
representations.  We leave a more detailed analysis of the 
structure of the triple $(A,{\cal H}, \hat D)$ for further 
developments.

\section{Comments and conclusions}  

{\em Spectral invariance.} We have disregarded gauge invariance.  
The action we have chosen is not a spectral invariant: it is not 
a function solely of the eigenvalues of the Dirac operators, as 
is the GR spectral action (\ref{actiongr}).  (The model we have 
chosen does not have enough degrees of freedom for accommodating 
gauges.)  Spectral invariance is crucial in GR, as, in 
particular, it incorporates diffeomorphism invariance.  The most 
interesting extension to the technique considered here should 
therefore, in our opinion, consists in the incorporation of 
spectral invariance.  In the case of a spectral invariant action, 
the eigenvalues $\lambda_{n}$ of the Dirac operators are natural 
real {\em gauge invariant\/} coordinates on $\cal C$.  We expect 
that in the quantum theory they be represented as self-adjoint 
operators $\hat\lambda_{n}$.  For GR, some information on the 
Poisson algebra of the $\lambda_{n}$'s was obtained in 
\cite{landi}.

{\em Zero point distance.} We may attach an intuitive physical 
interpretation to the ``zero point distance'' given by 
(\ref{main}) with $n=0$.  In the classical theory the two points 
$p$ and $p'$ can be at infinite distance (when $m=0$).  In non 
commutative geometry, this can be viewed as a ``classical'' 
limit.  In the quantum theory, on the other hand, there exist a 
maximal distance between $p$ and $p'$: the two ``sheets'' of the 
universe cannot separate in the quantum theory\footnote{I thank 
Thomas Sch\"ucker for pointing this out.}.

{\em Dimensions.} There are two natural possibilities of 
assigning physical dimensions to the Dirac operator in NCG. The 
traditional choice is to assign to the Dirac operator $D$ the 
dimension of a mass.  The second possibility is to assign it the 
dimension of an inverse length.  If $D$ is a mass, $G$ has 
dimensions $M/L$.  If $D$ is an inverse length, $G$ has 
dimensions $1/ML^{3}$.  If $D$ has the dimension of a mass, we 
must insert the Planck constant (or another constant with the 
same dimensions) in the formula (\ref{distance}) for the 
distance, in order to adjust the dimensions.  For instance we can 
write
\begin{equation}
	d(p,q) = sup_{\{a\in A, |[D,\pi(a)]|
	{\scriptscriptstyle \le} \hbar\}}\ |p(a)-q(a)|.  
	\label{distance3}
\end{equation}
This is not unreasonable, since in the commutative case the 
inequality is meant to fix the derivative of $a$ to be less than 
one, but the (commutative) Dirac operator with dimensions of a 
mass is proportional to $\hbar \frac{\partial}{\partial x}$.  On 
the other hand, it is a bit disturbing to invoke the Planck 
constant in a theory before quantizing it.  Equation 
(\ref{distance3}) yields, instead of (\ref{main})
\begin{equation} 
	d(p,p') 
    =  \frac{\hbar}{\sqrt{(2n+1) \hbar G }}
	= \frac{L_{P}}{\sqrt{2n+1}}, 
	\label{main3}
\end{equation}
where (\ref{LP}) is replaced by $ L_{P}=\sqrt{\frac{\hbar}{G}}$. 
If, instead, $D$ is an inverse length, all the equations in the 
text are dimensionally correct as they are.  The two ways of 
assigning dimension are obviously equivalent.  We find the second 
one ($D$ is an inverse length), which does not require us to 
insert the Planck constant in the definition of distance, more 
natural from the perspective of a gravitational theory.

{\em Functional quantization.} An alternative approach to the 
canonical quantization method we have employed here is Feynman's 
sum over histories approach.  To quantize our model a' la 
Feynman, we have to select a measure on the space of the 
histories, namely on ${\cal D}=C^{2}$ and integrate the 
exponential of the action with a source term added.  This defines 
the generating functional, from which the Green functions of the 
theory can be computed by derivation.  In a spectral invariant 
theory, this should yield an integration over the Dirac 
eigenvalues $\lambda_{n}$:
\begin{equation}
  Z = \int [d\lambda_{n}]\ e^{-i/\hbar S[D]} 
\end{equation}
Details on the possibility and the difficulties of this approach 
will be given elsewhere.

\vskip.5cm 
In conclusion, we have studied the quantum theory of an 
elementary dynamical noncommutative geometry.  We can derive a 
few lessons from this exercise.  First, a key object is the space 
${\cal D}_{(A,H)}$ of the Dirac operators compatible with given 
$A$ and $H$.  This is the kinematical arena of a dynamical 
noncommutative geometry.  A second key object is the subspace 
$\Gamma$ of ${\cal D}_{(A,H)}$ determined by the dynamical 
equations.  An important problem is the determination of the 
symplectic structure of $\Gamma$.  If a time evolution is 
defined, general techniques are available.  These extend to the 
case in which time evolution is a gauge as in GR \cite{landi}. In the model 
studied here, $\Gamma$ came equipped with a natural symplectic 
structure.  In the quantum theory, the matrix elements of $D$ are 
operators on a Hilbert space $K$.  The two Hilbert spaces $K$ and 
$H$ combine naturally into the Hilbert space $\cal H$, on which a 
quantum Dirac operator $\hat D$ is defined (see (\ref{hatD})).  
The quantum states of the geometry are given by irreducible 
representations of the algebra in $\cal H$.  The distance between 
two physical points (states over the algebra) which can take any 
nonnegative real value in the classical noncommutative geometry 
is quantized in the quantum theory, with a spectrum given in 
(\ref{qd}). 
The step from the simple model considered to a full theory 
including GR is obviously enormous.  The results described form 
just an exploration of the structures involved in quantizing a 
non-commutative-geometry.  Hopefully, these structures can be 
relevant in the quantization of more complete noncommutative 
models as well. 

\vskip.5cm 
\centerline{------------------------------------}
\vskip.5cm 

Great thanks to Abhay Ashtekar, for a very critical suggestion.  
To Thomas Krajewski, whom I have pestered with innumerable 
questions.  To Thomas Sch\"ucker and Bruno Iochum for carefully 
reading and criticizing the manuscript.  To Daniel Testard, 
Daniel Kastler, and everybody in the NCG group of the CPT for 
help, for listening, and for the wonderful welcome I have 
received in Marseille.  This work has been partially supported by 
NSF Grant PHY-95-15506.

\end{document}